# Model compression for faster structural separation of macromolecules captured by Cellular Electron Cryo-Tomography

Jialiang Guo[1*], Bo Zhou[2*], Xiangrui Zeng[3], Zachary Freyberg[4], and Min Xu[3**]

[1] Xi'an Jiaotong University, Xi'an, China
[2] Robotics Institute, Carnegie Mellon University, Pittsburgh, USA
[3] Computational Biology Department, Carnegie Mellon University, Pittsburgh, USA
[4] Departments of Psychiatry and Cell Biology, University of Pittsburgh, Pittsburgh, USA

**Abstract.** Electron Cryo-Tomography (ECT) enables 3D visualization of macromolecule structure inside single cells. Macromolecule classification approaches based on convolutional neural networks (CNN) were developed to separate millions of macromolecules captured from ECT systematically. However, given the fast accumulation of ECT data, it will soon become necessary to use CNN models to efficiently and accurately separate substantially more macromolecules at the prediction stage, which requires additional computational costs. To speed up the prediction, we compress classification models into compact neural networks with little in accuracy for deployment. Specifically, we propose to perform model compression through knowledge distillation. Firstly, a complex teacher network is trained to generate soft labels with better classification feasibility followed by training of customized student networks with simple architectures using the soft label to compress model complexity. Our tests demonstrate that our compressed models significantly reduce the number of parameters and time cost while maintaining similar classification accuracy.

## 1 Introduction and Related Work

Maintenance of cellular homeostasis is largely based upon the coordinated activities of an array of intracellular macromolecular complexes. The structural and functional roles of many of these complexes, to date, have been inferred from in vitro studies, often from purified samples. While such studies are extremely useful, the ability to study macromolecular complexes within their native cellular contexts may be even more informative. At present, live cell imaging does not have sufficient resolution to visualize the structure or distribution of individual macromolecular complexes within cells. Cellular Electron Cryo-Tomography (CECT), on the other hand, has the resolution to potentially visualize individual macromolecular complexes in three dimensions within cells preserved in a near-native state[6]. However, the systematic structural identification and recovery of the macromolecules captured by CECT is difficult due to the structural complexity and imaging limits in CECT tomograms. First, a key contributor of this complexity is the makeup of these large macromolecular structures which are composed of numerous components with highly dynamic conformations and interactions, thus limiting the overall complex resolution. Secondly, imaging limits of CECT such as low signal-to-noise ratio (SNR) and missing data effects (i.e. missing wedge) have further complicated systematic macromolecular structural recovery [8]. Therefore, it is critical to develop effective computational approaches to separate and average huge numbers (at least millions) of highly structurally diverse macromolecules represented by subtomograms (a *subtomogram* is a cubic sub-image that is likely to contain only one macromolecule). Recently, we have developed deep learning-based macromolecular structure classification approaches [8,2] with high discriminative ability and scalability to automatically process subtomograms. According to this methodology, convolutional neural networks (CNN) equipped with 3D filters were used to extract features from subtomograms and separate them into structurally homogeneous subgroups.

---

[*] Contributed equally
[**] Corresponding author

In order to achieve a significantly increased particle separation accuracy, it is necessary to search for a CNN model that has more layers and larger capacities. This is especially relevant given the rapid advances in the automation of CECT data acquisition, where efficient and accurate separation of tens of millions of macromolecules captured by CECT will soon become a new computational bottleneck towards achieving significantly improved systematic macromolecular structural identification and recovery. However, implementation of such a model would incur substantially increased computational and (GPU memory) storage costs compared with previously proven models. Indeed, in testing CNN models on subtomograms with a size of $40^3$ voxels and a voxel spacing of 0.92nm[8], on a computer with a single GPU, the training of previous CNN models using one million of these subtomograms for 20 epochs took more than 19 hours, and the separation of one million subtomograms could take up to 2 hours. Therefore, to substantially reduce computation and storage costs, it is critical to compress the CNN models into smaller ones with fewer parameters while maintaining the same accuracy, once highly accurate CNN models are obtained.

Use of compressed models has many notable advantages compared to the above models, including less prediction time, better deployment to other datasets, and often higher generalization ability [5]. Much previous work has been done to develop methods for model compression[4,7,1]. However, these methods were only tested on 2D images. To our knowledge, to date, little is known about the performance of model compression on CNN models designed for 3D image classification. In this paper, we will focus on reducing the complexity of deep neural networks by knowledge distillation. Among the preexisting models [2], we chose the DSRF3D-v2 (Deep Small Receptive Field) model for compression on the basis of considerations for processing time and performance. Corresponding student models are simpler versions of the original model. Knowledge in the teacher networks is compressed by generating soft labels which contain more information than the original labeled data. The student networks are trained with the soft labels. Our experiments show that with acceptable loss in classification accuracy, the size of distilled models can be significantly reduced. The reduction in the number of layers, and the simplification of structure significantly reduces the number of total parameters and processing time. Among the three student models we proposed, DSRF3D-v2-s1 achieved the best performance, which only requires approximately 1/20 of the original parameters, half of the original prediction time, but only loses 3% of accuracy as compared to the complex teacher model.

## 2 Method

In our previous work[8], a deep learning approach was proposed to subdivide structurally highly heterogeneous subtomograms into structurally more homogeneous smaller subsets through supervised feature extraction using CNN. Subtomograms are 3D grayscale sub-images representing particles within a cell, denoted as a function $f : \mathbb{R}^3 \to \mathbb{R}$. Each input $f$ of our CNN model is attached with a class label $\mathbf{y} := (y_1, \cdots, y_L)$, where $y_i \in \{0, 1\}, \sum_i y_i = 1$, and $L$ is the number of possible classes. The CNN outputs a probability vector $\mathbf{p} := (p_1, \cdots, p_L)$, where $p_i \in [0, 1]$ and predict the inputs $f$ to be in class $\arg\max_i p_i$. To achieve higher accuracy, we may still need to complicate the structure of our models. On the other hand, our testing data can contain millions of subtomograms and the image size can significantly increase under a higher resolution. Thus, it is necessary to find models with simpler structure for deployment.

### 2.1 Knowledge Distillation

Based on features extracted by prior convolutional and pooling layers, the last hidden layer in a convolutional neural network outputs logits $z_i$, which indicate the likelihood for each class and are converted into probabilities $p_i$ by the softmax function using equation 1.

$$p_i = \frac{\exp\{z_i/T\}}{\sum_j \exp\{z_j/T\}} \qquad (1)$$

The function compares each logit with others and uses the temperature $T$ to control the relative sizes of the output probabilities. The knowledge distillation technique[5] asserts that with more information than the original hard labels, the class probabilities produced by large neural networks can be used to guide small models to generalize information from the training data in the same way as large models. However, the probability vector tends to include many values near zero and only one value very close to 1; according to [5], the information that resides in the ratios of very small probabilities has very little influence on the cross-entropy cost function when training small models because the probabilities are too small. To recover the information learned by large models, [1] directly used logits may be employed as soft labels to transfer knowledge. The more generalized method knowledge distillation generates a soft target distribution as the soft label for each training case by using a high temperature in the softmax layer.

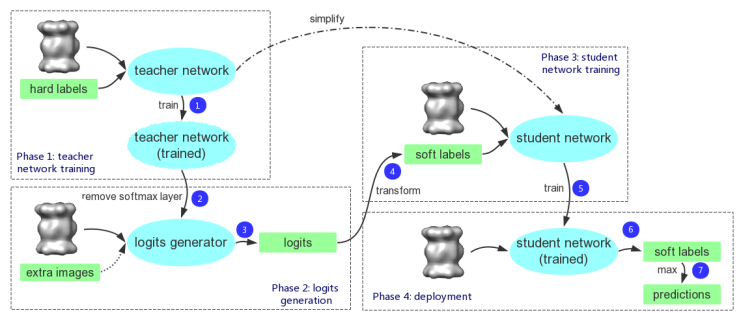

Fig. 1: Knowledge distillation flowchart for the deep-learning based subtomogram classification. The algorithm takes labeled tomographic data and a complex structure for the teacher network as inputs, and outputs a trained compact subtomogram classifier

Figure 1 illustrates the procedure of knowledge distillation for Deep Neural Networks that separates subtomograms extracted from CECT data. Phase 1 is regular supervised DNN training with labeled 3D images on the teacher network. The rather complex network structure is capable of finding a proper way to generalize information through training, with the knowledge of the learned function mostly stored in the logit layer. In Phase 2, we remove the softmax layer of the trained teacher network and obtain a DNN that outputs logits of the input subtomograms. Notably, since we do not need hard labels to generate logits, we can extend the training set with unlabeled images, if available.[5] The yielded logits are converted into soft labels in Phase 3 by the softmax function with a proper high temperature. The softened probabilities are then attached to the corresponding image to train the student network, typically a simplified version of the teacher network. [5] has explained that since the soft labels have high entropy, they provide much more information per training case than hard labels. Additionally, they possess much less variance in the gradient between training cases, so the small model can often be trained on much less data than the original, more cumbersome model. Furthermore, the compact student network uses a much higher learning rate but retain accuracy. In the deployment stage, the compact student network predicts the input subtomogram to be in the category corresponding to the max value in the output soft label.

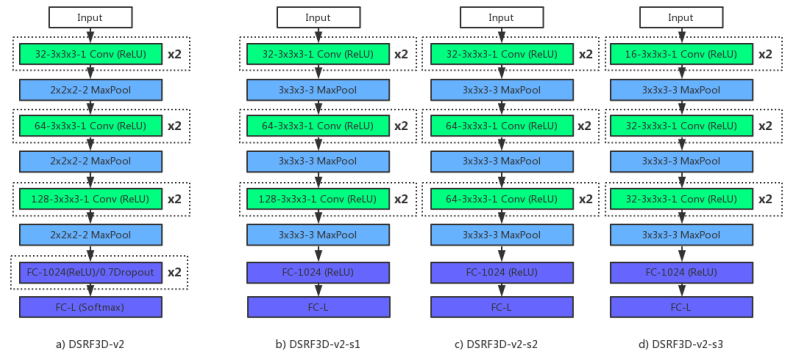

Fig. 2: The architectures of the teacher network and student networks. Each box with type and configuration corresponds to a layer in the neural network. For example, '32-3x3x3-1 Conv' is a convolutional layer with 32 3x3x3 filters and a stride of 1.'3x3x3-3 MaxPool' is a max pooling layer of a size of 3x3x3 and a stride of 3. 'FC-L' is a full connected layer with L(number of possible classes) neurons. The 'ReLU' and 'Softmax' in the bracket denote the type of activation function (Linear activation if not mentioned). "x2" outside the box means two sequential layers with the same configuration.

### 2.2 Teacher Network

The previous models in [2] have achieved a high accuracy of over 90% at separation of macromolecules extracted from CECT images. However, these models are relatively big in terms of the total number of parameters. Though the three models, DSRF3D-v2, RB3D and CB3D, are appropriate to serve as the prototypes of the teacher network, considering that the training process of CB3D takes too long and the performance of RB3D is relatively poor, we have opted to use DSRF3D-v2 as the teacher network in our compression.

As shown in figure 2, connected to the input layer are three sequential sets of stacked layers, each set consisting of two 3x3x3 3D convolutional layers and one 2x2x2 3D max pooling layer. Then it is followed by two fully connected layers with 70% dropout. All hidden layers are activated by ReLU. The final fully connected output layer has the same number of neurons as the number of possible structure class and uses softmax activation for outputs.

### 2.3 Student Networks

The student network is a simpler version of the teacher network. To simplify the teacher model, we chose to reduce the number of layers. However, directly connecting the first convolutional layer and pooling layer to the full connected layer brings more parameters due to the overwhelming number of local features in the first convolution operation. Therefore, the compression is implemented by simplifying the convolutional layers, pooling layers and eliminating one of the two fully connected layers. Essentially, the training samples with soft labels brings easier knowledge for the CNN to learn. The student network with 'weaker' layers has sufficient capability for the classification task, as the information in the soft labels distill the knowledge about how the teacher network learned from the dataset. To weaken convolutional layers, the simplest way is to reduce the number of filters. Increasing the pooling size of pooling layers makes the pooling operation performed on a greater region and ignores some local features.

We demonstrate three different student network configurations in Figure 2, which we believe have best maintained the accuracy while efficiently compressing the model. The student networks share similar structure with the teacher network, except that the dropout layers and one fully connected layer with 1024 neurons were dismissed. Based on that, details of the additional student networks' modifications are listed as follows:

a) In DSRF3D-v2-s1, the pooling size and stride of all the three max pooling layers are increased from 2x2x2 to 3x3x3.

b) In DSRF3D-v2-s2, besides performing modification in a), the last two convolutional layers drop half of original 128 filters.

c) In DSRF3D-v2-s3, besides performing modification in a), the number of filters in the six convolutional layers change from (32,32,64,64,128,128) to (16,16,32,32,32,32).

These modifications lead to a significant drop in the total number of parameters. Removal of one FC-1024 at the end of the network reduces approximately 1 million parameters. Increasing the stride of the pooling layer from 2x2x2 to 3x3x3 has decreased the overall number of parameters by nearly 70% . Applying the modification to all three pooling layers has resulted in a decrease of 97%. Halving the number of filters in a convolutional layer halves the overall number of parameter of this layer.

## 3 Implementation Details

We implemented this work by using Keras with Tensorflow as a back-end. The experiments are performed on a computer with three Nvidia GTX 1080 GPUs, one Intel Core i7-6800K CPU and 128GB memory.

The input of our models is labeled subtomogram, which is the volumetric image data extracted from CECT data. In [2], 12 datasets of simulated subtomograms were generated with different levels of SNR and tilt angle ranges. In this work, we chose simulated subtomograms with a tilt angle range of $\pm 60°$ and a SNR level equal to 0.05 as our training data, which is further split into 14720 training samples, 3680 validation samples and 4600 test samples.

## 4 Experimental Results

In this paper, we are using three student models to distill knowledge and compare them with the original uncompressed model. Our student networks are trained with the same configuration as in our previous work[2]. We use a stochastic gradient descent (SGD) optimizer and minimize the categorical cross-entropy cost function by adding Nesterov momentum of 0.9. The initial learning rate is set at 0.005 with a decay factor of 1e-7. The training processes are performed with a batch size of 64 for 20 epochs and will stop early if the classification performance shows no improvement over 5 consecutive epochs based on the loss function. The temperature in the softmax function is set to 5 heuristically by experience.

Table 1: COMPARISON BETWEEN TEACHER AND STUDENTS

| model | #parameter | accuracy[1] | accuracy [2] | test time (s) |
| --- | --- | --- | --- | --- |
| DSRF3D-v2 | 18,316,599 | 95.89% | / | 15.91 |
| DSRF3D-v2-s1 | 1,014,071 | 85.65% | 93.82% | 9.41 |
| DSRF3D-v2-s2 | 506,039 | 84.26% | 91.17% | 8.95 |
| DSRF3D-v2-s3 | 161,639 | 72.17% | 89.83% | 5.58 |

[1] test accuracy when the model is trained with hard labels
[2] test accuracy when the model is trained with soft labels

As shown in Table 1, the teacher network is significantly compressed with a remarkable decrease in the total number of parameters. In assaying test time, our results indicates a significant reduction in test duration, demonstrating a great potential for deployment. On the other hand, the classification accuracy of three student network suffers little loss. The DSRF3D-v2-s1 maintains almost the same accuracy, with the next two models at an accuracy of approximately 90%. An important fact we should note is that if we use the original hard labels to train our student networks, the test accuracy is relatively unsatisfactory and significantly lower than the accuracy when the model is trained with soft labels, which contain information from the more capable teacher network. This improvement in accuracy, to some degree, provides proof that the student networks themselves are not strong enough in structure to extract sufficient information from the training data and that the knowledge distilled from the teacher network accounts for the high accuracy of the student networks.

Table 2: EVALUATION OF DIFFERENT STUDENT NETWORKS

| model | compression rate[1] | accuracy rate[2] | speedup[3] |
| --- | --- | --- | --- |
| DSRF3D-v2 | 1 | 1 | 1 |
| DSRF3D-v2-s1 | 18.1 | 0.978 | 1.69 |
| DSRF3D-v2-s2 | 36.2 | 0.951 | 1.78 |
| DSRF3D-v2-s3 | 113.3 | 0.937 | 2.85 |

[1] the ratio of number of parameters
[2] the ratio of test accuracy
[3] the ratio of test time (all ratios are of the teacher network to model of interest)

Table 2 gives three evaluation metrics[3] of the student networks. Naturally, as the compression rate and the speedup increase, the accuracy preserved in the student model gradually decreases. Therefore, when simplifying the teacher network, we must prevent the layers from being oversimplified so that the obtained student network has enough capability of imitating the teacher network. In practice, we must consider the trade-off between loss of accuracy and parameter compression. Like the student networks demonstrated above, choosing proper student network architectures that have fewer parameters but that remain within the range of acceptable accuracy loss is critical.

## 5 Conclusion

In this paper, we have proposed a model compression approach for CECT data based on knowledge distillation to compress previously proposed high-accuracy deep neural network models for subtomogram classification. With DSRF3D-v2 as the teacher network, we have correspondingly designed relatively compact models serving as student networks, among which DSRF3D-v2-s1 achieves the best classification accuracy. Typically, a higher compression rate will result in a greater loss of accuracy. In contrast, our distilled models significantly reduce the number of parameters and processing time cost, while keeping almost the same classification accuracy. Therefore, computation cost and storage cost are effectively reduced, making it possible to deploy our neural networks in practice to classify massive datasets of CECT images, especially those with a large image size.

## 6 Acknowledgements

This work was supported in part by U.S. National Institutes of Health (NIH) grant P41 GM103712. M.X acknowledges support of Samuel and Emma Winters Foundation.